\documentclass[12pt]{iopart}
\usepackage{iopams}
\usepackage{epsfig}
\newcommand{\erfc}{\mathrm{erfc}}
\begin{document}
\title{Evolution in random fitness landscapes: 
the infinite sites model}
\author{Su-Chan Park and Joachim Krug}
\address{Institut f\"ur Theoretische Physik, Universit\"at zu K\"oln, Z\"ulpicher Str. 77, 50937 K\"oln, Germany}
\eads{\mailto{psc@thp.uni-koeln.de} and \mailto{krug@thp.uni-koeln.de}}
\date{today}
\begin{abstract}
We consider the evolution of an asexually reproducing population in an uncorrelated random fitness
landscape in the limit of infinite genome size, which implies that each mutation generates a 
new fitness value drawn from a probability distribution $g(w)$. This is the finite population
version of Kingman's house of cards model [J.F.C. Kingman, \textit{J. Appl. Probab.}
\textbf{15}, 1 (1978)]. In contrast to Kingman's work, the focus here is on unbounded
distributions $g(w)$ which lead to an indefinite growth of the population fitness. 
The model is solved analytically in the limit
of infinite population size $N \to \infty$ and simulated numerically for finite $N$. When the genome-wide
mutation probability $U$ is small, the long time behavior of the model reduces to a point process
of fixation events, which is referred to as a \textit{diluted record process} (DRP).
The DRP is similar to the standard record process except that a 
new record candidate (a number that exceeds all previous entries in the sequence) 
is accepted only with a certain probability that depends on the 
values of the current record and the candidate.
We develop a systematic analytic approximation scheme for the DRP. 
At finite $U$ the fitness frequency distribution of the population decomposes into a stationary
part due to mutations and a traveling wave component due to selection, which 
is shown to imply a reduction
of the mean fitness by a factor of $1-U$ compared to the $U \to 0$ limit.  

\end{abstract}
\maketitle
\section{Introduction}
A fruitful exchange of concepts and methods has taken place between evolutionary population 
biology and the statistical physics of disordered systems over the past several decades
\cite{Anderson1983,Stein1992,Lassig2002}. On the most basic level, one imagines that a biological 
population evolves by searching a high-dimensional landscape for fitness peaks, in much
the same way as a disordered system relaxes towards its low energy configurations
\cite{Ebeling1984,Nattermann1989}. Not surprisingly, extremal statistics arguments
play a prominent role in both contexts 
\cite{Nattermann1989,Bovier2006,Kauffman1987,Kauffman1993,Sibani1998,Orr2005,Krug2005}.
To make the analogy more precise, we note that 
the inheritable characters of an individual (its \textit{genotype}) 
are encoded in a genetic sequence
(consisting of nucleotide letters or the alleles of genes), which for many purposes
can be reduced to a binary sequence $\sigma = (\sigma_1,...,\sigma_L)$ 
of fixed length $L$. For a statistical physicist it is very natural to assign
the values $\sigma_i = \pm 1$ to the letters, and to treat the sequence as,
e.g., a row of spins in the two-dimensional Ising model \cite{Leuthausser1987}
or a configuration of a quantum spin chain \cite{Baake1997}. A 
\textit{fitness landscape} is then a real-valued function $W(\sigma)$ on the 
$L$-dimensional sequence space, analogous to the (negative) energy of the
spin system. 

The notion of a fitness landscape is a venerable and persistent image in evolutionary
biology \cite{Gavrilets2004,JK2006}, but it has always been plagued by a 
certain elusiveness, in the sense that very little is known about the
fitness landscapes in which real organisms evolve. This situation may eventually change, 
as the experimental mapping of genotypic fitness becomes
feasible for simple microbial systems \cite{Weinreich2005,Poelwijk2007}. Meanwhile it is reasonable
to handle our ignorance of real fitness landscapes by treating $W(\sigma)$ as 
a realization of a suitably chosen ensemble of random functions. This approach was
pioneered by Kauffman and coworkers \cite{Kauffman1987,Kauffman1993}, who introduced 
the NK family of random 
fitness landscapes which are closely analogous to Derrida's $p$-spin model of spin
glasses developed a few years earlier \cite{Bovier2006,Derrida1981}. Two limiting cases of the
model are of interest here: The \textit{random energy model} (REM), in which
fitness (or energy) is assigned randomly without correlations
to the genotypes (or spin configurations), and the case in which the letters
$\sigma_i$ contribute independently (\textit{multiplicatively} for
discrete time dynamics \cite{JK2006}) to the fitness. 
In the latter case there is
always a single fitness maximum, which explains why this is also referred to as the
\textit{Fujiyama} landscape. In the evolutionary context deviations from 
multiplicative fitness are associated with \textit{epistasis} \cite{Weinreich2005}.
Within the NK family, the REM landscape is maximally epistatic \cite{WW2005}. 

In previous work the evolutionary process in the REM landscape has been studied
mostly in the limit of infinite population size, where fluctuations due to sampling noise
(also known as \textit{genetic drift} in population genetics) are ignored 
\cite{JK2006}. While this allows one to derive a
rather complete picture of both stationary \cite{Franz1993,Franz1997}  
and time-dependent  
\cite{Krug2003,Jain2005,Sire2006,Jain2007} properties of the model, 
the assumption of an infinite population is unrealistic, because
the number of possible genotypes $2^L$ exceeds any conceivable population
size $N$ already for moderate values of $L$. On the other hand, 
individual-based simulation studies which explicitly follow the population
through sequence space are restricted to rather short sequences \cite{Amitrano1989,JK2007}. 

In this contribution we therefore propose to perform the limit of 
infinite sequence length at finite population size $N$. 
This kind of limit is well known 
in population genetics, where it is viewed alternatively as a
limit on the number of genetic loci or sites (the sequence length $L$ in our setting)
or as a limit on the number of alleles (the number of possible values that the
variables $\sigma_i$ can take). Although mathematically distinct, the two 
variants are equivalent for our purposes. 

The implementation of the infinite
sites limit for the REM landscape is straightforward: For $L \to \infty$
every mutation leads to a new genotype, whose fitness can be randomly generated
without need to keep track of the neighborhood relations of the sequence space.
In this way large populations can be simulated efficiently for many generations.
Moreover, we show that for long times and small mutation rates the model reduces to a simple
point process which is partly tractable analytically. We also present an analytic
solution of the infinite population limit of the model, which is useful for 
describing the evolution of finite populations at early times and provides
important insights into the structure of the fitness distribution.  

The infinite population version of 
our model was introduced by Kingman in 1978 \cite{Kingman1978} and is known as the 
\textit{house of cards model} \cite{Burger2000}. 
This term refers to the idea that the genetic organization of an organism 
is very fragile, such that it is completely disrupted 
by any change in the genome which therefore leads to a new fitness that
is uncorrelated with the fitness of the parent genotype.  
Previous studies of the finite
population dynamics of this model have been concerned mostly with the
regime of weak selection, where a balance is established between
deleterious and beneficial mutations  \cite{OT1990,T1991,G1994}.
In contrast, in the present paper we consider the strong selection regime
where the population fitness increases without bound (see \sref{Sec:Fixation} for a 
definition of these regimes). 

We give a brief outline of the paper. In the next section we introduce the basic
Wright-Fisher dynamics for asexually reproducing populations of constant
size $N$, and describe how to implement the infinite sites limit. 
The evolution of the fitness distribution in the deterministic limit
$N \to \infty$ is discussed in \sref{Sec:infiniteN}. In \sref{Sec:IF}
we turn to the long-time behavior of finite populations. The key observation     
is that (for unbounded fitness distributions)
\textit{fixation events} in which a favorable mutation spreads in the
population become rare and well-separated as the mean fitness increases.
The evolution then reduces to a simple point process closely related to the
dynamics of records \cite{Sibani1998,Krug2005,Glick1978,Arnold1998}. We show that to leading order
the mean population fitness achieved up to time $t$ is given by the mutation of largest
fitness that has been encountered,  
and we develop an analytic framework to compute sub-leading contributions    
to the mean fitness and the fitness variance.  
Finally, conclusions and some open problems are presented in \sref{Sec:Conc}. 

\section{Wright-Fisher dynamics in the infinite sites limit}
\begin{figure}[t]
\centerline{
\includegraphics[width=0.6\textwidth]{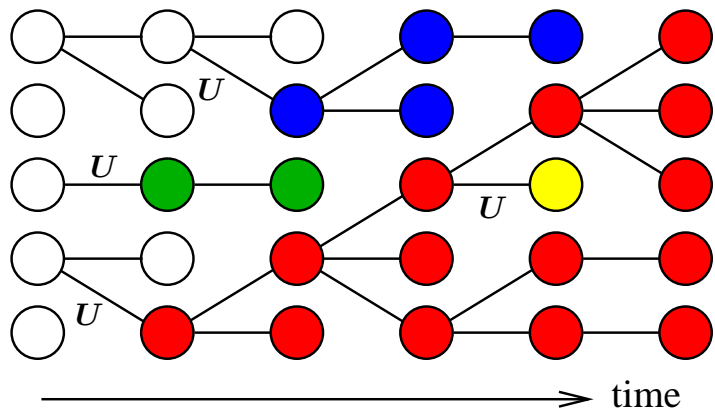}
}
\caption{\label{Fig:WF} Schematic of the WF model explained in the text.
The population size is fixed at $N=5$.
Time direction is indicated by the arrow (5 generations). Circles signify the individuals and 
different colors mean different genotypes which arise by mutation
with probability $U$. The vertical 
location of individuals has no significance (the population is assumed
to be well-mixed, without spatial structure). 
Initially, all individuals have the same genotype (hence the same color). Line segments
connecting individuals show who begets whom. Each individual can have only one parent but a parent
can have many offspring. 
After five generations, the red mutation which arose in a single individual
at time 2 is \textit{fixed} in the population 
(see \sref{Sec:Fixation} for further
discussion of the fixation process).
}
\end{figure}
We consider a population of $N$ individuals reproducing asexually,
in discrete, non-overlapping generations. The basic Wright-Fisher
(WF) dynamics \cite{Wright1931,Fisher1930} of evolution can be described as follows. 
Each individual $i$ is assigned fitness $w_{i,t}$ ($i=1,
\ldots,N$) at generation $t$. Initially, all individuals have the same 
genotype and accordingly the same fitness $w_{i,0} = 1$.
At every generation, all individuals are replaced. The probability
that a new individual is an offspring of the parent $i$ is proportional to 
parent's fitness $w_{i,t}$. To be more accurate, this probability is 
$w_{i,t}/(  \bar w_t N)$, where $\bar w_t = \sum_{i=1}^N w_{i,t}/N$
is the mean fitness\footnote{In this paper, 
we use three different notations for `mean fitness'. First,
$\bar w_t$ is a random variable defined as the population 
average of the fitness in a specific 
realization of the WF model.  Second, we use $\langle \bar w_t \rangle$
to denote the average of $\bar w_t$ over independent realizations;
for infinite populations, this distinction is unnecessary. Third, 
by $\bar w(\tau)$ we mean $\langle \bar w_t \rangle$ at $t = \tau /(NU)$
for given $N$ and $U$ 
to emphasize the similarity between the simulation studies and
the continuous time point process introduced in \sref{Sec:IF}.} at generation $t$. 
In the actual simulation, we do not 
discern different progenitors if they have the same genotype. 
Instead, the number of progeny of a given genotype
is determined from the multinomial distribution
with a probability also proportional to the population of that genotype.
The multinomial distributed numbers are chosen by sampling 
correlated binomial random numbers; see, e.g., \cite{Park2007,D1986}.
Since this reproduction scheme is invariant under the
multiplication of a constant to the fitness of all individuals, setting 
the average initial fitness to unity implies no loss of generality.
Once all individuals are replaced,
a mutation can change the genotype of each one with probability $U$. 
A cartoon illustrating the WF model is depicted in \fref{Fig:WF}.

In the infinite sites limit, every genotype occurring by mutations can appear
only once (there are no \textit{recurrent} mutations \cite{KO2005}).  
In the REM landscape the fitness 
of the mutant is a random number $w$ drawn from some fixed
distribution $g(w)$, independent of the number of sites affected by the mutation as 
well as of the parental genotype. By contrast, the multiplicative fitness landscape
is represented in the infinite sites limit by drawing a random 
selection coefficient $s$ and generating the new fitness $w'$
from the parental fitness $w$ by multiplication, $w' = (1+s) w$
\cite{Park2007,GL1998,O2000,W2004}.

The model is completely specified by the population size $N$, the mutation probability
$U$ and the choice of the mutation distribution $g(w)$.  
To make contact with the evolutionary dynamics in a finite space of sequences
of length $L$ \cite{JK2007}, we note that $U$ is the probability that \textit{at least
one mutation occurs in a genotype}. We therefore have the relation
\begin{equation}
\label{U} 
U = 1 - (1-\mu)^L \simeq 1 - \exp(-\mu L), 
\end{equation}
where $\mu$ is the mutation probability
per site, and the limit $\mu \to 0$, $L \to \infty$ is implied.
Typical values for $U$ range from 0.0025 for \textit{E. coli} to 
0.15 for humans and 0.9985 for the bacteriophage $Q \beta$ \cite{JK2006}.

\begin{figure}[t]
\includegraphics[width=\textwidth]{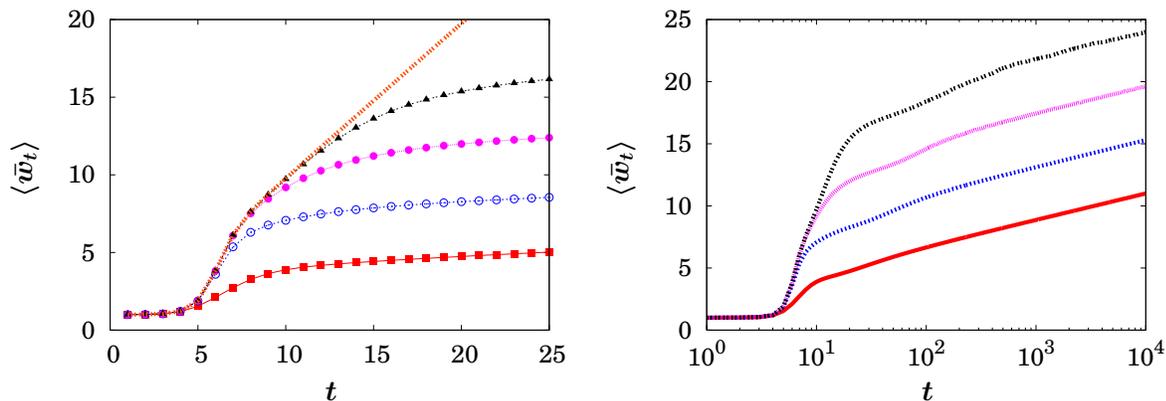}
\caption{\label{Fig:Nmu} Mean fitness of the WF model for finite and infinite
populations. For finite populations, the mean fitness is also averaged over independent
runs. The mutation probability $U$ is set to 0.01 and 
the mutation distribution is $g(w) = e^{-w}$. From bottom to top,
the population size $N$ increases ($N = 10^3$, $10^5$, $10^7$, and $10^9$).
The number of independent runs is
$8\times 10^6 \; (N=10^3)$, $8 \times 10^5 \; (N=10^5)$, $20~000 \; (N=10^7)$, and $4000 \; (N=10^9)$.
Left panel shows the short time simulation results and compares them
to the infinite population calculation in \sref{Sec:infiniteN}.
Right panel depicts log-linear plots of the mean fitness
for finite populations which clearly shows a logarithmic increase of the 
mean fitness at long times. 
}
\end{figure}

An important difference between the models with finite and infinite genome size is that
in the latter case there are no local fitness maxima in which populations can get
trapped when $U$ is small \cite{Kauffman1987,Kauffman1993,WW2005,JK2007}. Hence we expect an indefinite increase
of the population mean fitness $\bar w_t $ when the fitness distribution $g$ is unbounded.
In the following sections we explore how the behavior of $\bar w_t$ depends on the model parameters
and the choice of $g$. Some representative results from simulations with different values of 
$N$ and $U$ and the exponential mutation distribution 
\begin{equation}
\label{expdist}
g(w) = e^{-w} 
\end{equation}
are shown in \fref{Fig:Nmu}. The majority of our results were obtained with the choice
(\ref{expdist}).

\section{\label{Sec:infiniteN}Infinite populations}

In this section, we consider the infinite population dynamics in the REM fitness
landscape with an infinite number of sites. 
In this paper, we take the infinite site
$L \to \infty$ limit first, followed by the infinite population limit $N \to \infty$;
note that these two limiting procedures do not obviously commute.

\subsection{Calculation of the mean fitness}

Let $ f_t(w) dw$ be the fraction of individuals in the population with a fitness between $w$ and $w+dw$.
In the limit of infinite population size this evolves deterministically according to \cite{Kingman1978}  
\begin{equation}
\label{recursion}
 f_{t+1}(w) = (1 - U) \frac{w  f_t(w)}{\bar w_t}  +U   g(w),
\end{equation}
where $\bar w_t = \int_0^\infty dw \; w  f_t(w)$ is the mean fitness.
If $U = 1$, trivially and intuitively $f_t(w) \equiv g(w)$.
The nonlinearity can be removed by introducing \cite{JK2006,Kingman1978}
$h_t(w) = f_t(w) \prod_{k=0}^{t-1} \bar w_k$ with $h_0(w) = f_0(w)$
which satisfies
\begin{equation}
h_{t+1}(w) 
= ( 1 - U) w h_t(w) + U g(w) X_{t+1},
\end{equation}
where
\begin{equation}
\label{Eq:Self_con}
X_{t+1} = \prod_{k=0}^{t} \bar w_k = \int_0^\infty dw \; h_{t+1}(w).
\end{equation}
The last equality is due to the fact that $f_t(w)$ remains normalized.
Formally, we can solve the above equation such that
\begin{equation}
\frac{h_t(w)}{(1-U)^t }
=f_0(w) w^t + U g(w) \sum_{k = 1}^t \frac{w^{t-k} X_k}{(1-U)^k}.
\end{equation}
From the self-consistency condition \eref{Eq:Self_con}, one can find
a recursion relation for $X_t$,
\begin{equation}
Y_t = \phi_t + U \sum_{k = 1}^t Y_k G_{t-k}, 
\end{equation}
where $Y_t = X_t/(1-U)^t$, 
and the quantities
\begin{equation}
\label{phit}
\phi_t = \int_0^\infty dw \; f_0(w) w^t, \;\;\;\; 
G_t = \int_0^\infty dw \; g(w) w^t
\end{equation}
are moments of the initial condition $f_0(w)$ and the fitness distribution $g(w)$,
respectively. 
The recursion relation for $Y_t$ can be rewritten in the form
\begin{equation}
Y_t = \frac{\phi_t}{1-U} + \frac{U}{1-U} \sum_{k = 1}^{t-1} Y_k G_{t-k},
\label{Eq:Iteration}
\end{equation}
which is (at least numerically) solvable by iteration.
Once we have calculated $Y_t$, the mean fitness follows naturally from
\begin{equation}
\bar w_t = (1-U)\frac{Y_{t+1}}{Y_t}.
\end{equation} 

Although we cannot find a complete analytic expression for $\bar w_t$,
the leading asymptotic behavior can be easily found for some cases.
If, for large $t$, 
\begin{equation}
\label{criterion}
G_{t-1} \gg \phi_t \;\; \textrm{and} \;\; G_t \gg G_{k} G_{t - k} \;\;
\textrm{for} \; \textrm{all} \;\; 1 \le k \le t-1, 
\end{equation}
one can say with small error that
\begin{equation}
Y_t \simeq \frac{U}{1-U} Y_1 G_{t-1} \;\;
\Rightarrow \;\; \bar w_t \simeq (1-U) \frac{G_t}{G_{t-1}} .
\label{Eq:gamma_result}
\end{equation}
The criterion (\ref{criterion}) is actually not very restrictive. If
$g(w)$ has the form of a stretched exponential multiplied
by a power law,
\begin{equation}
g(w) = \frac{\beta}{w_0 \Gamma\left(\frac{\nu+1}{\beta} \right ) }\left ( \frac{w}{w_0}\right )^{\nu} \exp\left ( - \left( \frac{w}{w_0} \right)^\beta \right ),
\label{Eq:g_dist}
\end{equation}
where $w_0$ is a constant and $\Gamma(x)$ is the gamma function,
the $t$-th moment of $g(w)$ becomes
\begin{equation}
G_t = w_0^t \Gamma\left ( \frac{\nu + t +1}{\beta} \right )
\Gamma\left(\frac{\nu+1}{\beta} \right )^{-1},
\label{Eq:Gen_g}
\end{equation}
which satisfies (\ref{criterion}). 
Using Stirling's formula $\Gamma(z) \sim z^z e^{-z} \sqrt{2 \pi/z}$,
one finds that
\begin{equation}
\label{Eq:Inf_Asym}
\frac{\bar w_t}{w_0(1-U)}\approx \left ( \frac{\nu + t + 1}{\beta} \right)^{1/\beta}
\sim \left(\frac{t}{\beta}\right )^{\frac{1}{\beta}},
\end{equation}
a result that was also reported by Kingman \cite{Kingman1978}. 

\begin{figure}[t]
\centerline{
\includegraphics[width=0.6\textwidth]{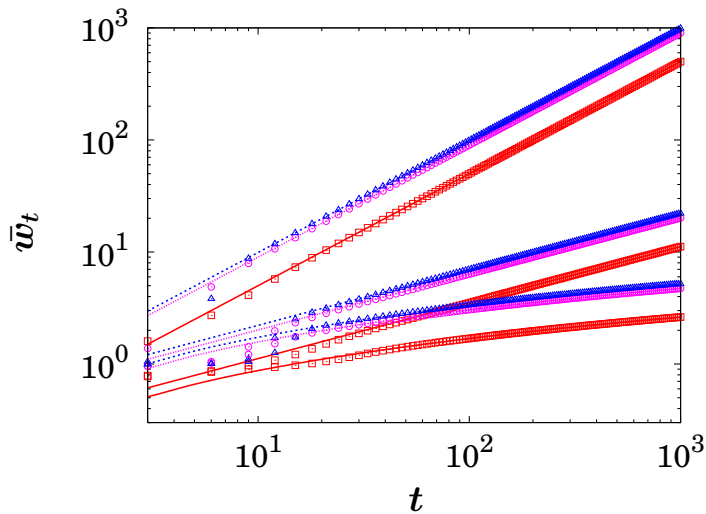}
}
\caption{\label{Fig:inf_fit} Log-log plots of mean fitness in the infinite
population limit as a function of generation $t$ for exponential 
[$\nu=0$ and $\beta=1$ in \eref{Eq:g_dist}; upper three data sets],
Gaussian [$\nu=0$ and $\beta = 2$ in \eref{Eq:g_dist}; middle three data sets] 
and Gumbel-type distributions 
\eref{Eq:Gum_Dis} (lower three data sets). For all data sets,
$w_0$ and $f_0(w)$ are set to 1 and $\delta(w-1)$, respectively. 
For each case, the mutation probability is $U = 0.5$ (red), $U = 0.1$ (purple), 
and $U = 0.01$ (blue) from bottom to top.
The lines are the approximate solutions \eref{Eq:Inf_Asym} and \eref{Eq:Gum_Asym} with
the respective parameters.}
\end{figure}
When $\beta \rightarrow \infty$, 
one might expect that the power law (\ref{Eq:Inf_Asym}) turns into a
logarithmic increase of the fitness. As an example of an unbounded distribution
that decays more rapidly than (\ref{Eq:g_dist}), we consider 
in Appendix A the Gumbel-type mutation distribution
\begin{equation}
g(w) = \frac{e}{w_0} \exp\left ( \frac{w}{w_0} - e^{w/w_0} \right ).
\label{Eq:Gum_Dis}
\end{equation}
\Fref{Fig:inf_fit} 
compares the exact behavior obtained by
iterating \eref{Eq:Iteration} to the asymptotic expression~\eref{Eq:Inf_Asym}
and \eref{Eq:Gum_Asym} for the exponential, Gaussian, and Gumbel-type distributions. 
In this comparison,
the initial fitness distribution is set to
$f_0(w) = \delta(w-1)$ and $w_0 = 1$.
The analytic approximations are seen to be very accurate in all cases, with
error less than 1 \% after 100 generations.

Up to now, we have assumed that $G_t$ is finite for any $t$. If $g(w)$
has a power law tail 
\begin{equation}
\label{powerg}
g(w) =  \alpha (1+w)^{-(\alpha+1)}, 
\end{equation}
however, $G_t$ becomes infinite at $t = t_0 = \{\alpha\}$ where $\{x\}$ means
the smallest integer not smaller than $x$.
This means that $Y_{t_0}$ is finite but $Y_{t_0 +1}$ is infinite, i.e.
$\bar w_{t}$ becomes infinite at $t=t_0$. This peculiarity of the deterministic
selection dynamics has also been observed in the multiplicative case~\cite{Park2007},
and it is consistent with the behavior of the infinite population dynamics in 
a finite sequence space, where the population reaches the global fitness maximum
in a single time step for large $L$ \cite{Jain2005}. 

For mutation distributions $g(w)$ with bounded support the recursion (\ref{recursion})
approaches a limiting distribution $f_\infty(w)$ which has been described in detail
by Kingman \cite{Kingman1978}. Remarkably, for certain choices of $g(w)$ one finds
a condensation phenomenon in which $f_\infty(w)$ develops a $\delta$-function
singularity at the maximally possible fitness $w_\mathrm{max}$ (set by the upper limit of the support
of $g$ or of $f_0$, whichever is larger). The asymptotic mean fitness is bounded
from below by $(1 - U) w_\mathrm{max}$. In the remainder of the paper we will restrict
the discussion to unbounded mutation distributions.

\subsection{Fitness distribution}

Note that the mutation 
rate (provided it is nonzero) enters (\ref{Eq:Inf_Asym}) only through the prefactor
$1-U$, which is known in population genetics as the \textit{mutational load}
\cite{Burger2000}. Its origin can be traced back to the fact that, in each generation,
the fitness is reset randomly for a fraction $U$ of the population, so that selection
can act only on the remaining fraction $1-U$ [compare to (\ref{recursion})]. 

To clarify this effect in a quantitative manner, let us find out what the frequency distribution 
looks like in the asymptotic regime. To this end, we calculate higher moments of 
$f_t(w)$ from \eref{recursion} along with \eref{Eq:gamma_result}. For convenience, let
\begin{equation}
\zeta_n(t) \equiv \frac{1}{(1-U) w_0^n } \int w^n f_t(w) dw.
\end{equation}
A recursion relation for the moments can be found 
by multiplying $w^{n-1}$ to both sides of \eref{recursion} and integrating
over $w$, which yields
\begin{equation}
\zeta_n(t) =\zeta_1(t) \left ( \zeta_{n-1}(t+1) - \frac{U}{1-U}\frac{G_{n-1}}{w_0^{n-1}}   
\right ),
\end{equation}
and $\zeta_1(t)$ can be read off from \eref{Eq:gamma_result} or \eref{Eq:Inf_Asym}.
The formal solution for $\zeta_n(t)$ is
\begin{equation}
\zeta_n(t) = \prod_{k=0}^{n-1} \zeta_1(t+k) - \frac{U}{1-U} \sum_{k=1}^{n-1} 
\frac{G_{k}}{w_0^k} \prod_{\ell=0}^{n-k-1} \zeta_1(t+\ell).
\end{equation}
Since $w_0 \zeta_1(t)  \approx G_{t}/G_{t-1}$ for large $t$, $\zeta_n(t)$ becomes
\begin{equation}
(1-U)w_0^n \zeta_n(t) \approx  U G_n + (1-U)\frac{G_{t+n-1}}{G_{t-1}} - U \sum_{k=1}^{n} 
\frac{G_k G_{t+n-1-k}}{G_{t-1}} .
\label{Eq:zeta_n}
\end{equation}
Constructing the Laplace transform (or the moment generating function) of the 
frequency distribution using \eref{Eq:zeta_n} such that
\begin{equation}
\int e^{-z w} f_t(w) dw \approx U \tilde g(z) + (1 - U  ) \tilde \varphi_t(z) - U \tilde \psi_t(z),
\label{Eq:Laplace}
\end{equation}
where 
\begin{equation}
\tilde g(z) = \sum_{n=0}^\infty \frac{(-z)^n}{n!} G_n
= \int dw e^{-zw} g(w)
\end{equation}
is the Laplace transform of $g(w)$ and 
\begin{equation}
\tilde \varphi_t(z) = \sum_{n=0}^\infty \frac{(-z)^n}{n!} \frac{G_{t+n-1}}{G_{t-1}},\quad
\tilde \psi_t(z) = \sum_{n=1}^\infty \frac{(-z)^n}{n!} \sum_{k=1}^{n}
\frac{G_k G_{t+n-1-k}}{G_{t-1}},
\end{equation}
we can in turn find the frequency distribution through
inverse Laplace transformation. 
Since the Laplace transformation is linear, $f_t(w)$ can be written as
\begin{equation}
f_t(w)\approx  U g(w) + (1-U) T_t(w) - U  \Psi_t(w),
\end{equation}
where $T_t(w)$ and $\Psi_t(w)$ are the inverse Laplace transformations of $\tilde \varphi_t(z)$
and $\tilde \psi_t(z)$, respectively.
Since  ($n> 1$)
\begin{equation}
G_n \ll  \sum_{k=1}^n \frac{G_k G_{t+n-1-k}}{G_{t-1}}
\ll \frac{G_{t+n-1}}{G_{t-1}}
\end{equation}
because of the criterion \eref{criterion}, we expect
that when $w$ is small [large], the dominant contribution for $f_t(w)$ comes from
$g(w)$ [$T_t(w)$]. Hence we neglect the contribution from $\Psi_t(w)$, which gives
\begin{equation}
f_t(w) \approx U g(w) + (1-U) T_t(w).
\label{decomp}
\end{equation}

For the exponential distribution ($\nu=0,~\beta=1$), the analytic
form of $\tilde \varphi(z)$ can be found.
For this case, $G_t = w_0^t t!$, $\tilde g(z) = (1+z w_0)^{-1}$, and
\begin{equation}
\tilde \varphi_t(z) = \sum_{n=0}^\infty \frac{(-z)^n}{n!} \frac{G_{t+n-1}}{G_{t-1}}
= \sum_{n=0}^\infty {t+n-1 \choose n} (-z w_0)^n = (1 + z w_0)^{-t}.
\end{equation}
Hence the frequency distribution is
\begin{equation}
f_t(w) \approx \frac{U}{w_0} e^{-\frac{w}{w_0}} + \frac{1-U}{w_0 (t-1)!} 
  \left ( \frac{w}{w_0} \right )^{t-1} 
e^{-\frac{w}{w_0}}.
\label{Eq:fre_dis}
\end{equation}
One can easily check that the mean and the variance at large $t$ become
\footnote{In \cite{Kingman1978}, $\delta w^2$ was also calculated but 
the factor $(1-U)$ is missing in the result.}
\begin{equation}
\bar w_t \approx (1-U) w_0 t,\quad
\delta w^2 \approx U (1-U) w_0^2 t^2,
\label{Eq:full_var}
\end{equation}
where 
$\delta w_t^2 = \int w^2 f_t(w) dw - \bar w_t^2 $.
Standard deviation and mean are of the same order and
$\delta w$ is even larger than $\bar w_t$ if $U>\frac{1}{2}$. The
origin of such a large spread is the division
of (\ref{Eq:fre_dis}) into two widely separated distributions:
A time-independent part, arising from the mutations, 
and a traveling wave reflecting the selection dynamics. 
Then the mean and the variance of $T_t(w)$ in the asymptotic regime are
found to be 
$t w_0$ and $t w_0^2$, respectively. Hence the spread of $T_t$ is much smaller
than the mean in the asymptotic regimes, which cannot be
appreciated from the full variance in \eref{Eq:full_var}.
By the central limit theorem\footnote{Recall that the Gamma distribution with
integer $t$ can be interpreted as the distribution of
the sum of $t$ independent and identically distributed random variables
with exponential distribution~\cite{FellerII}.}, $T_t(w)$ is approximated by the Gaussian distribution
\begin{equation}
T_t(w) \approx \frac{1}{\sqrt{2 \pi t w_0^2}} \exp\left ( - \frac{(w - w_0 t)^2}{2 tw_0^2 } \right ).
\label{Eq:Gauss_app}
\end{equation}

Although we cannot find an analytic form of $f_t(w)$ for the general class
of distributions (\ref{Eq:g_dist}), 
the qualitative form is expected to be the same as in the exponential
case, that is, a superposition of a 
stationary distribution due to mutations and a Gaussian travelling wave.
Based on this conjecture, we can approximate the travelling wave for the general case.
From the generating
function  of $T_t(w)$ which is $\tilde \varphi(z) $,
one can get the leading behavior of the mean and variance of $T_t(w)$ such as 
\begin{eqnarray}
-\tilde \varphi'(0) = \frac{G_t}{G_{t-1}} \approx \left ( \frac{t}{\beta}
\right )^{\frac{1}{\beta}},\\
\tilde \varphi''(0) - \tilde \varphi'(0)^2= \frac{G_{t+1}}{G_{t-1}} - 
\left (\frac{G_{t}}{G_{t-1}} \right )^2
\approx \frac{1}{\beta^2}
\left ( \frac{t}{\beta} \right)^{\frac{2-\beta}{\beta}},
\end{eqnarray}
that is, the travelling wave is Gaussian with mean $\bar w_t/(1-U)$ and
width $\sim t^{(2-\beta)/(2 \beta)}$.

\begin{figure}[t]
\centerline{
\includegraphics[width=\textwidth]{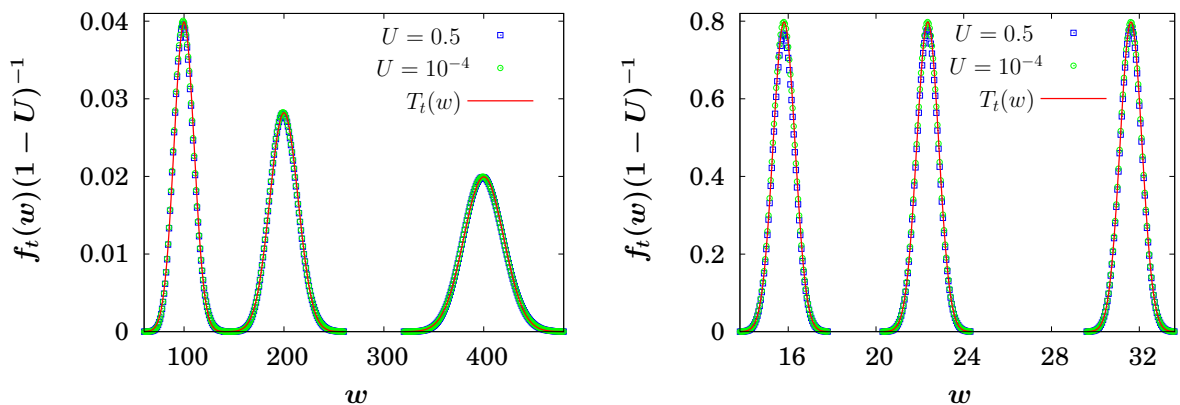}
}
\caption{\label{Fig:travel} Comparison of the exact numerical solution for
the frequency distribution with the travelling wave equation.
Left Panel: Frequency distributions at generation $100$, $200$, and $400$ (from left
to right) are shown for $U=0.5$ (square) and $U=10^{-4}$ (circle).
There is a slight mismatch due to the neglect of $\Psi_t(w)$ but 
ingeneral the travelling wave solution approximate the true
distribution quite well. Right panel : Similar study to the Left panel
with the Gaussian distribution. The data are collected at generation
$500$, $1000$, and $2000$. The Gaussian approximation is almost perfect.
The frequency distribution for small $w$ is $U g(w)$ for both
panels as reasoned in the text (data not shown).}
\end{figure}

Since we know the exact values of $\bar w_t$ from the numerical iteration of 
\eref{Eq:Iteration}, we can also numerically calculate the frequency density
from \eref{recursion}.
\Fref{Fig:travel} numerically confirms for the exponential and Gaussian mutation distribution
that the frequency distribution
is divided by two parts and the travelling part takes the Gaussian form with the
predicted mean and variance in this section.
The decomposition \eref{decomp} will play a considerable role in 
understanding the behavior of finite
populations with large $U$, see \sref{Sec:finiteU}.

\section{\label{Sec:IF}Finite populations}
\subsection{Fixation and clonal interference}

\label{Sec:Fixation}

An important element in the evolution of finite populations is the process of 
\textit{fixation}, in which a mutation that is initially present in a single individual
spreads in the population and eventually is shared by all individuals
(see \fref{Fig:WF} for illustration). 
Consider the simple case of a single mutant of fitness $w'$ entering a 
genetically homogeneous population in which all individuals have the same fitness
$w$. The success of the mutant is determined by the selection coefficient
\begin{equation}
\label{selection} 
s = \frac{w'}{w} -1, 
\end{equation}
which is positive (negative) for beneficial (deleterious)
mutations. For the WF model the fixation probability is given
approximately by \cite{K1962}
\begin{equation}
\label{pigen}
\pi_N(s) \approx \frac{1 - e^{-2s}}{1 - e^{-2Ns}}.
\end{equation}
When selection is strong, in the sense that 
\begin{equation}
\label{strong}
N \vert s \vert \gg 1,
\end{equation}
it can be seen from (\ref{pigen}) that 
fixation of deleterious mutations becomes exponentially unlikely, while 
a beneficial mutation is fixed with probability  
\begin{equation}
\label{pi}
\pi(s) \approx 1 - e^{-2s}.
\end{equation}
Previous work on the house of cards model in finite populations has
been concerned with the weak selection regime, where $N \vert s \vert \approx 1$
\cite{OT1990,T1991,G1994}. This regime can be realized by choosing a 
mutation distribution
$g(w)$ whose standard deviation is much smaller than the mean.
However, in the context of the present paper the
strong selection criterion (\ref{strong}) is always satisfied. 

The mean time to fixation of a beneficial 
mutation is given by \cite{K1962} 
\begin{equation}
\label{tfix}
t_\mathrm{fix} \approx \frac{\ln N}{s}.
\end{equation}
Different evolutionary regimes arise from the comparison of 
$t_\mathrm{fix}$ to the expected time interval between 
the emergence of beneficial mutations that are destined for fixation.
Denoting by $U_b$ the beneficial mutation probability per individual,
beneficial mutations arise at rate $N U_b$. A mutation fixes with probability
$\pi(s_b) \approx 2 s_b$, where $s_b$ is the typical selection coefficient
which is assumed to be small, so that (\ref{pi}) can be approximated by
$2 s$. Then the waiting time between fixation events is 
$t_\mathrm{mut} \approx 1/(2 N U_b s_b)$. 
When $t_\mathrm{fix} \ll t_\mathrm{mut}$ or \cite{W2004} 
\begin{equation}
\label{clonal}
2 N \ln N U_b \ll 1,
\end{equation} 
beneficial mutations arise rarely and fix independently,
a regime that is referred to as \textit{periodic selection}
\cite{deVisser2006}. In the opposite case 
$2 N \ln N U_b \gg 1$ clones originating from different
mutants compete for fixation, a phenomenon that 
is known as \textit{clonal interference} \cite{GL1998}.

In previous studies of clonal interference
\cite{Park2007,GL1998,O2000,W2004,deVisser2006} it has usually been assumed
that $U_b$ is a constant parameter, in which case
(\ref{clonal}) is a condition on the population size $N$ which
is violated when $N$ becomes large. However, in a rugged landscape
the supply of beneficial mutations decreases as the mean fitness grows.
If the fitness distribution
of the population is well clustered around its mean $\bar w$, the 
probability of beneficial mutations can be estimated by
\begin{equation}
\label{Ub}
U_b(\bar w) = U \; 
\mathrm{Prob}[w > \bar w] = U \int_{\bar w}^\infty dw \; g(w),
\end{equation}
which vanishes for $\bar w \to \infty$ for any unbounded distribution
$g(w)$. Thus clonal interference is a \textit{transient} phenomenon
in rugged fitness landscapes. Asymptotically almost all mutations are deleterious,
and hence $U$ can be identified with the probability of deleterious mutations. 

\subsection{Instantaneous fixation and the diluted record process}

The fact that the criterion \eref{clonal} is asymptotically satisfied
for unbounded $g(w)$ implies that the fixation of beneficial 
mutations occurs as independent events
that can then as well be treated as instantaneous ($t_\mathrm{fix} \to 0$).
Deleterious mutations do not fix, but they lower the mean population fitness
through the mutational load $1 - U$ (compare to \sref{Sec:infiniteN}). For the time
being,
we will neglect this effect, which amounts to taking $U \to 0$.
We will show in \sref{Sec:finiteU} how the influence of deleterious mutations
can be approximately reinstated.  

When the effects of deleterious mutations are ignored,
the population fitness between fixation events is equal to the
fitness of the last beneficial mutation that was successfully fixed,
and the population is genetically homogeneous at all times (except
for the instances of fixation). 
The model reduces to a simple point process that can be 
informally described as follows \cite{T1991,G1994}: 

\begin{itemize}

\item[(i)] Mutations with fitness values drawn
randomly and independently from $g(w)$ are generated 
in discrete time according to a 
Poisson distribution with mean $N U$.

\item[(ii)] The fitness of the mutant
is compared to the current population fitness; 
deleterious mutations are discarded while beneficial mutations 
are fixed with probability $\pi(s)$ given by (\ref{pi}). 

\item[(iii)] The fitness of 
the successfully fixed mutant replaces the current population fitness, which
therefore evolves according to a piecewise constant, strictly increasing jump process.

\end{itemize}
If every beneficial mutation were fixed, such that $\pi(s) = \Theta(s)$, the 
process described above would be identical to a variant 
of the well-known problem of record statistics for sequences
of independent, identically distributed variables \cite{Glick1978,Arnold1998},
in which a Poisson-distributed number of new variables is created in each
discrete time step. When $\pi(s) < 1$ some of the record events are lost in 
a way that is correlated to the corresponding record values. The 
process defined by the rules (i)-(iii) is referred to in the following 
as the \textit{diluted record process} (DRP), and it will be studied
extensively in the next three subsections. 

\begin{figure}[t]
\includegraphics[width=\textwidth]{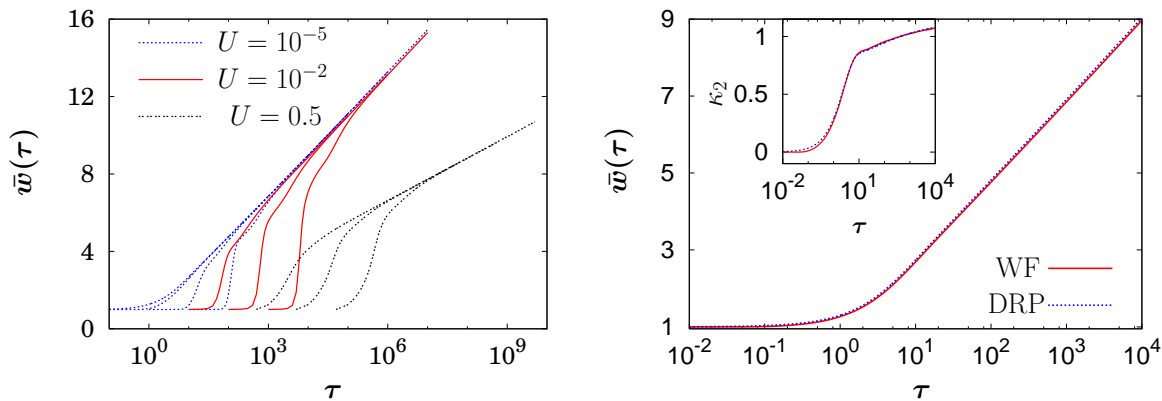}
\caption{\label{Fig:WFS} Left panel: Semilogarithmic plots of the fitness vs
dimensionless time $\tau = N U t$ for $U = 10^{-5}$ ($N=10^3,\;10^4,\;
10^5,\;10^6$),  $U=10^{-2}$ and $U=0.5$ ($N=10^3,\;10^4,\; 10^5$) 
obtained from simulations of the WF model.
The mutation distribution and the initial frequency distribution are same
as those in \fref{Fig:Nmu}. 
Right panel: The comparison of the simulation of the WF model
to the DRP for $N = 10^3$ and $U = 10^{-5}$ ($\tau = 10^{-2} t$).
The difference in the fitness is barely observable. Inset: Same type of the comparison
for the variance $\kappa_2$ [see \eref{Eq:meancor_master}] of the fitness.}
\end{figure}

For long times, when beneficial mutations become increasingly rare, the 
discrete unit of time is unimportant and the dependence on the  
system parameters $N$ and $U$ can be eliminated by using the dimensionless
time variable $\tau = U N t$. Asymptotically the DRP is therefore fully specified by the 
functions $\pi(s)$ and $g(w)$.  
A comparison between the DRP and the full WF dynamics
is shown in \fref{Fig:WFS}. For large $N$ an initial regime can be identified in which
clonal interference reduces the fitness in the WF model compared to the DRP, 
and for large $U$ the fitness is reduced by the mutational load
effect, but for long times and $U \ll 1$ the agreement is seen to be essentially perfect.
In this and the following three subsections, we restrict ourselves to $U \ll 1$ and 
the analysis of the behavior for large $U$ is deferred to \sref{Sec:finiteU}.

Clearly the record process (RP) provides an upper bound on the DRP at all times \cite{G1994};
for completeness, a derivation of the fitness distribution for the RP is provided in 
Appendix B. We will now argue
that the RP bound is in fact saturated for the leading order behavior of the mean fitness,
in the sense that 
\begin{equation}
\label{bound}
\lim_{t \to \infty} \frac{\langle w \rangle_\mathrm{DRP}}{\langle w \rangle_\mathrm{RP}} = 1.
\end{equation}
To see why this is so, denote by $w_1$ the largest fitness value that has appeared up to time
$\tau$, and by $w_2$ the largest fitness that has also been fixed. Then $w(\tau) = w_1$ in 
the record process and $w(\tau) = w_2$ in the DRP. If $w_1$ were asymptotically larger
than $w_2$ in the sense that $w_1/w_2 \geq C > 1$, then the selection coefficient of 
$w_1$ in a background of $w_2$ is $s_{12} = w_1/w_2 - 1 \geq C-1 > 0$ and the corresponding 
fixation probability $\pi(s_{12})$ is bounded away from zero. It follows that $w_1$ is fixed
with finite probability, in contradiction to our assumption that $w_2$ is the largest fitness
that has been fixed. 

We will see later that the average fitness $w_2$ at time $\tau$ for
the exponential distribution is of order $w_2 \sim \ln \tau - \ln \ln \tau$,
while $w_1 \sim \ln \tau + \mathrm{const}$.
As the difference between the largest and $k$th largest value
is of order $\ln k$ for exponential random variables \cite{D1970}, this implies that
the rank of $w_2$ among the $\tau$ fitness values that have been created
up to time $\tau$ is $O(\ln \tau)$.

In the following subsections we will develop some analytic tools to systematically
compute the mean fitness and higher fitness moments for the DRP.

\subsection{Mean field approximation}

In the mean field approximation (MFA) the fitness distribution of the population is 
characterized only by its mean, which will be denoted by
$m(\tau)$ in the following. The probability that a new mutation with 
arbitrary fitness $w' > m$ is fixed is given by
\begin{equation}
p_\mathrm{fix} = \int_{m}^\infty \pi \left( \frac{w'-m}{m} \right )
 g(w') dw' = m \int_0^\infty
\pi(x) g(m x+m) dx, 
\end{equation}
and the waiting time until this happens (in dimensionless units)
is $\Delta \tau = 1/p_\mathrm{fix}$. 
Once fixation occurs, the population fitness increases by the amount
\begin{equation}
\Delta w = \frac{\int_{m}^\infty w' \pi\left( \frac{w'-m}{m} \right )
  g(w') dw'}{\int_{m}^\infty \pi\left( \frac{w'-m}{m} \right )
 g(w') dw'} 
- m
= m \frac{\int_0^\infty x \pi(x) g(m x+m) dx}{\int_0^\infty \pi(x) g(m x+m) dx},
\end{equation}
and $m(\tau)$ is obtained by solving the differential equation 
\begin{equation}
\label{ODE}
\frac{dm}{d\tau} = \frac{\Delta w}{\Delta \tau} = p_\mathrm{fix}\Delta w .
\end{equation}
For the exponential distribution (\ref{expdist}) this takes the explicit form
\begin{equation}
\frac{dm}{d\tau} = 4 \frac{m+1}{(m+2)^2} e^{-m},
\label{Eq:mean_exp}
\end{equation}
with the solution
\begin{equation}
e^{m} ( m + 2 ) + I(m) =  4( \tau - \tau_0) + e^{w_0} ( w_0 + 2 ) + I(w_0),
\label{Eq:fitnessEq}
\end{equation}
where 
\begin{figure}[t]
\centerline{
\includegraphics[width=0.7\textwidth]{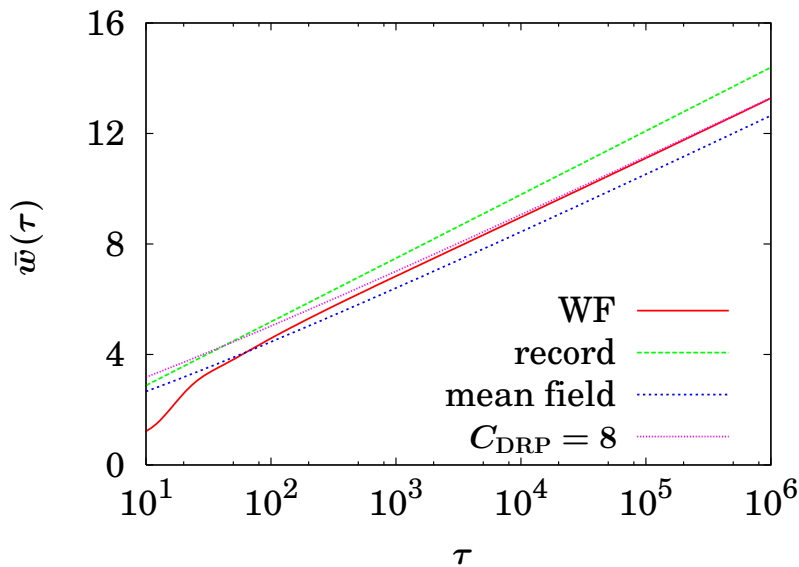}
}
\caption{\label{Fig:comall} Comparison of the WF simulation with the record,
mean field, and the improved approximation scheme in \sref{Sec:fluc} with the
stable value of $C_\mathrm{DRP} = 8$.
The WF simulation results are obtained with the exponential 
mutation distribution (\ref{expdist}) with $N=10^5$ and $U = 10^{-5}$ 
($\tau = t$). As explained in the text, mean field theory and record dynamics give
lower and upper bounds on the true asymptotics.
}
\end{figure}
\begin{equation}
I(m) \equiv \int_0^m \frac{e^x}{x+1} dx \sim \sum_{n=0}^\infty n! \left (
\frac{e^m}{(m+1)^{n+1}} - 1 \right ).
\end{equation}
The asymptotic expansion which is a divergent series is 
obtained by integration by parts.
Now assume that $\tau$ is extremely large, then the approximate inversion formula
of \eref{Eq:fitnessEq} will be the solution of the equation
\begin{equation}
m e^m \approx 4 \tau \;\;\; \Rightarrow 
\;\;\;\; m + \ln m \approx \ln  (4 \tau).
\label{Eq:fitapp}
\end{equation}
From \eref{Eq:fitapp}, one can easily see that
$\lim_{\tau \rightarrow \infty} \ln\tau / m = 1$. 
Let $m(\tau) = ( 1 + f(\tau)) \ln (4 \tau)$ where $f(\tau) \rightarrow 0$ as 
$\tau \rightarrow \infty$. Then
\begin{equation}
f(\tau) \ln (4 \tau) + \ln \ln (4 \tau) + \ln (1 + f(\tau)) = 0 \;\;\;
\rightarrow 
\;\;\; f(\tau) \approx - \frac{\ln \ln 4 \tau }{1 + \ln 4 \tau},
\label{Eq:app}
\end{equation}
which, as expected, approaches zero in the asymptotic regime.
According to extreme value statistics, 
the largest fitness up to time $\tau$ is of order
$\ln \tau + \gamma$, where $\gamma (\approx 0.5772)$ is the Euler number. 
But the leading behavior of $m(\tau)$ is $\ln \tau
+ \ln 4$, which seems contradictory. This apparent paradox  
can be resolved by looking at
the difference $\ln (\tau) - m(\tau) = O(\ln \ln t) > 0$, 
which diverges for $\tau \to \infty$. 
In \fref{Fig:comall}, the MFA solution \eref{Eq:fitapp} with the 
correction \eref{Eq:app} is compared to the WF simulation and the record problem.

For fitness distributions
with a power law tail \eref{powerg}, we get ($\alpha>1$)
\begin{equation}
p_\mathrm{fix} = \alpha m^{-\alpha} J(\alpha+1,m), \quad
\Delta w = m \frac{J(\alpha,m) - J(\alpha+1,m)}{J(\alpha+1,m)}-1,
\end{equation}
where 
\begin{equation}
J(\alpha,m) = \int_1^\infty  \left (1 - e^{-2(x-1)} \right ) \left( \frac{1}{m} + x
\right )^{-\alpha} dx.
\end{equation}
To have a meaningful result, we should restrict ourselves to the case $\alpha >1$.
Hence \eref{ODE} for the power law case becomes
\begin{equation}
\eqalign{
\frac{dm}{d\tau} &= \alpha m^{-\alpha +1} ( J(\alpha,m) - J(\alpha+1,m) ) -\alpha  m^{-\alpha} J(\alpha+1,m)\\
&\approx \alpha m^{-\alpha +1} ( K(\alpha) - K(\alpha+1) ),
}
\label{Eq:mean_power}
\end{equation}
where $K(\alpha) = J(\alpha,\infty)$.
Evaluation of the differential equation~\eref{Eq:mean_power} in the asymptotic regime
($m \gg 1$) yields
\begin{equation}
m \approx \tau^{1/\alpha} ( \alpha^2 ( K(\alpha) - K(\alpha+1) )^{1/\alpha}
\equiv \tau^{1/\alpha} L(\alpha).
\label{Eq:mean_powerfactor}
\end{equation}
To complete the analysis, let us compare $L(\alpha)$ to the prefactor obtained 
in (\ref{Eq:power_rec}) for the record process.
When $\alpha -1 \ll 1$, the two prefactors have the expansion
\begin{equation}
\eqalign{
L(\alpha) = \frac{1}{\alpha-1}
+ \ln ( \alpha-1) + 0.916~014 + o(1),\\
\Gamma\left ( \frac{\alpha-1}{\alpha} \right ) = \frac{1}{\alpha-1} + (1 - \gamma)
+ o(1),
}
\end{equation}
which shows that the RP prefactor is larger than that of the 
MFA in this regime.
When $\alpha \gg 1$, the asymptotic behavior of the MFA prefactor
can be obtained by integration by parts. One finds
\begin{equation}
\eqalign{
L(\alpha) = 1 + \frac{1}{\alpha} \ln (4/\alpha)
+ o(1/\alpha)<1,\\
\Gamma\left (1-\frac{1}{\alpha} \right ) = 1 + \frac{\gamma}{\alpha} + o(1/\alpha)>1,
}
\end{equation}
which also suggests that the MFA prediction is smaller than the RP value.
Actually, the MFA prefactor 
becomes smaller than unity when $\alpha > 3.533~18$ while $\Gamma((\alpha-1)/\alpha)$
remains larger than unity.
In between, one can numerically check that the RP prefactor
is always larger than that of the MFA.
In fact, we will show in the next subsection that the MFA always provides a lower bound
on the true mean fitness of the DRP. 
To summarize the mean field theory for the power law distribution,
the MFA fitness is of the same order as the extremal
statistics estimate \eref{power}, but the smaller prefactor 
is not consistent with the relation \eref{bound}. 

\begin{figure}[t]
\centerline{
\includegraphics[width=0.8\textwidth]{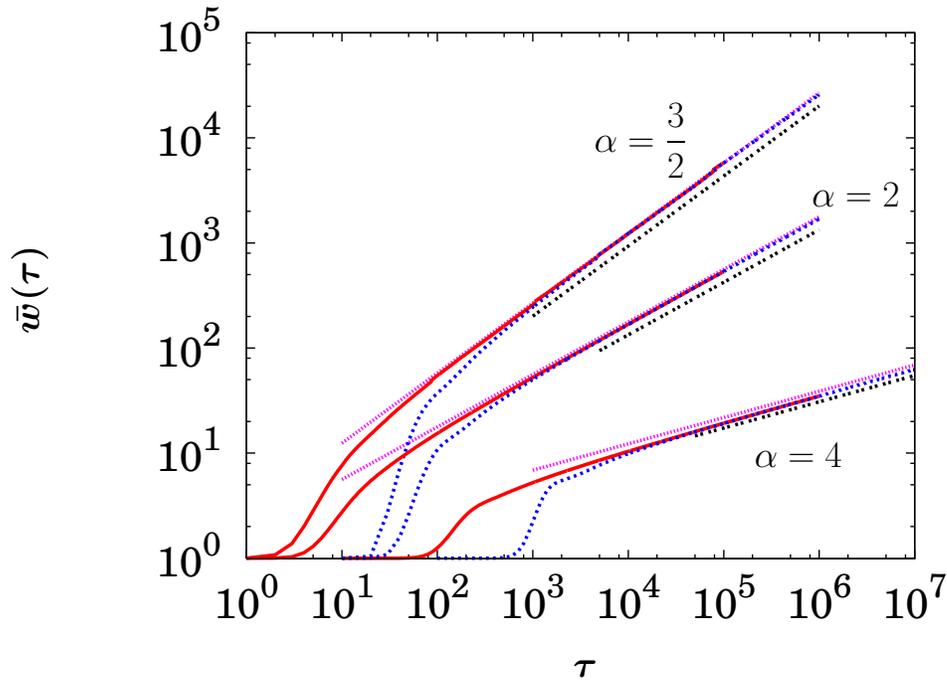}
}
\caption{\label{Fig:powerdis} Log-log plots of $\bar w (\tau)$ vs $\tau$ for the simulation
results of the WF dynamics
with the power law distribution \eref{powerg} with $\alpha = \frac{3}{2}$ (upper
data set), $\alpha = 2$ (middle data set), and $\alpha=4$ (lower data set).
The mutation probability 
is set to $U = 10^{-4}$ and the population sizes are $10^4$ and $10^5$ for $\alpha=\frac{3}{2}$
and 2, and $10^5$ and $10^6$ for $\alpha = 4$, respectively.
For comparison, the record solution 
\eref{Eq:power_rec} (straight line segments above the simulation results) and the 
MF prediction \eref{Eq:mean_powerfactor} (straight line segments below the simulation results) are also depicted.
}
\end{figure}
\Fref{Fig:powerdis} compares the WF simulation results with the record solution~\eref{Eq:power_rec}
and the MF result~\eref{Eq:mean_powerfactor} for three values of $\alpha$.
For $\alpha = \frac{3}{2}$, the RP result is in good agreement with the
WF simulations but as $\alpha$ increases, both bounds increasingly deviate from
the simulation data. Since for
$\alpha \rightarrow \infty$ the power law distribution approaches
a distribution of exponential type, the results for the exponential  
distribution summarized in \fref{Fig:comall} indicate
that a similar discrepancy should be expected 
for larger values of $\alpha$. For $\alpha\le 2$, the variance of the distribution
of record values is infinite, which implies that the value of a new record is usually
much larger than the previous record and $\pi(s) \approx 1$. 
In this case the DRP becomes identical to the RP. On the other hand, as
$\alpha$ gets bigger, the ratio of two consecutive records approaches unity in the
asymptotic regime. This implies stronger corrections to the leading 
behavior which, according to \eref{bound}, should still be given by the RP.

\subsection{\label{Sec:fluc}Master equation}

To improve on the mean field approximation, we will first
derive an equation for the transition probability $p(w,\tau|w_0,0)$
of the DRP~\cite{WW2005,T1991}. Since the model is a Markov process, the conditional probability
$p(w,t+dt|w',t')$ completely specifies the evolution equation. We have 
\begin{equation}
\eqalign{
p(w,\tau+d\tau|w',\tau)  
=& \delta(w-w') \left ( 1 - d\tau \int_{w'}^\infty \pi
\left (\frac{x-w'}{w'} \right )g(x) dx \right )\\
&+ d \tau \Theta(w-w')g(w) \pi\left ( \frac{w-w'}{w'} \right ) . }
\end{equation}
Using the Markov property, the equation for
$p(w,\tau)\equiv p(w,\tau|w_0,\tau_0)$ can be found as follows:
\begin{equation}
\fl
\eqalign{
p(w,\tau+d\tau) = \int dx \; p(w,\tau+d\tau|x,\tau)p(x,\tau)\\
 = p(w,\tau) \left ( 1 - d\tau \int_{w}^\infty \pi\left (\frac{x-w}{w}\right ) g(x) dx \right )
+ g(w) d \tau \int_{w_0}^w \pi \left( \frac{w-x}{x}\right ) p(x,\tau) dx,}
\end{equation}
and therefore\footnote{When the fixation of deleterious mutations is included by using
the expression (\ref{pigen}) for the fixation probability, for suitable choices
of $g(w)$ the master equation satisfies detailed
balance with respect to a stationary distribution \cite{T1991}. In the present setting
where selection is assumed to be strong in the sense of (\ref{strong}), this stationary
distribution cannot be reached on reasonable time scales.}
\begin{equation}
\fl
\frac{dp(w,\tau)}{d\tau} = 
-p(w,\tau) \int_{w}^\infty \pi\left (\frac{x-w}{w} \right ) g(x) dx + g(w) \int_{w_0}^w  \pi \left ( \frac{w-x}{x}
\right ) p(x,\tau) dx.
\label{Eq:master}
\end{equation}
The description by the master equation (\ref{Eq:master}) is appropriate for a Markov
process like the DRP which has non-continuous sample paths \cite{G1990}.
One can easily check that the record probability \eref{Eq:rec_gen} solves \eref{Eq:master}
when $\pi(x)=\Theta(x)$.

From (\ref{Eq:master}) evolution equations for arbitrary expectation values
$\langle f(w,\tau) \rangle $ can be obtained in the form
\begin{equation}
\fl
\eqalign{\frac{d}{d\tau} \langle f(w,\tau) \rangle \equiv \frac{d}{d\tau} \int dw \; p(w,\tau) f(w,\tau) =
\int dw \left ( f(w,\tau) \frac{\partial p(w,\tau)}{\partial \tau} + \frac{\partial f(w,\tau)}{\partial \tau}
p(w,\tau) \right )\\
 = \left \langle w \int_0^\infty \left [ f(wx+w,\tau) - f(w,\tau) \right ]\pi(x) g(wx+w) dx \right \rangle + \left \langle \frac{\partial f(w,\tau)}{\partial \tau} \right \rangle.}
\label{Eq:mean_time}
\end{equation}
For example, the equation for the centered normalized moment
$\kappa_n \equiv (w-\langle w \rangle )^n / n!$ is
\begin{eqnarray}
\label{Eq:meanfit_master}
\frac{d \langle w \rangle}{d \tau} = \left
\langle w^2 \int_0^\infty x \pi(x) g(wx + w)dx  \right \rangle ,\\
\frac{d \kappa_n}{d \tau} = 
\sum_{r=1}^{n} \left \langle \frac{w^{r+1}}{r!}  \frac{(\delta w)^{n-r}}{(n-r)!}
\int_0^\infty dx x^{r} \pi(x) g(wx + w) \right \rangle - \kappa_{n-1} \frac{d \langle w \rangle}{d \tau},
\label{Eq:meancor_master}
\end{eqnarray}
where $\delta w \equiv w - \langle w \rangle$.
For the cases of the exponential distribution (\ref{expdist}) and the power law \eref{powerg},
\eref{Eq:meanfit_master} becomes
\begin{eqnarray}
\label{Eq:RLP_1}
\frac{d \langle w \rangle}{d \tau} =
\left \langle \frac{4 (w + 1)}{(w +2 )^2} e^{-w} \right \rangle,\\
\frac{d \langle w \rangle}{d \tau} =
\alpha \left \langle w^{-\alpha +1} ( J(\alpha,w) - J(\alpha+1,w) ) - w^{-\alpha} J(\alpha+1,w)
\right \rangle,
\label{Eq:power_master}
\end{eqnarray}
which reduce to the mean field equations~\eref{Eq:mean_exp} and \eref{Eq:mean_power} if we approximate 
$\langle \chi(w) \rangle \approx \chi(\langle w \rangle)$, where $\chi(w)$ is the function inside
the brackets on the right hand side of (\ref{Eq:RLP_1},\ref{Eq:power_master}).
Again note that \eref{Eq:power_master} is meaningful only if
$\alpha >1$.
Since $\chi(w)$ is a convex function asymptotically, that is, $\chi''(w)>0$ for $w\gg 1$, 
$\chi(\langle w \rangle ) \le \langle \chi(w) \rangle$, 
which means that the mean field theory yields a \textit{lower}
bound on the true asymptotics.

\subsection{Moment expansion}
We now develop a systematic approximation scheme which extends beyond mean field
theory. First we write down the differential equations for $m \equiv \langle w \rangle$,
$\kappa_2,\ldots,\kappa_\ell$ which are available from Eqs.~\eref{Eq:meanfit_master} and
\eref{Eq:meancor_master} once $g(w)$ is given. Then we expand the terms on the right 
hand side, of the general form $\langle \chi(w) \rangle$, up to $\kappa_\ell$ in such a way that
\begin{equation}
\langle \chi (m + \delta w) \rangle = \chi(m) + \kappa_2 \frac{ d^2 \chi(m)}{d m^2} + 
\kappa_3 \frac{d^3 \chi(m)}{d m^3} + \ldots
\end{equation}
and keep terms only up to $\kappa_\ell$ in all equations for $\kappa_n$ ($\ell\ge n$).
If we only keep terms up to $\ell=1$, then we arrive at the mean field equation.
If we keep terms up to $\ell=2$, then we have
\begin{equation}
\eqalign{
\frac{dm}{d\tau} &= \frac{4(m+1)}{(m+2)^2 e^m} + \frac{4(m^3 + 7 m^2 + 14 m + 2)}{(2+m)^4 e^m} \kappa_2,\\
\frac{d \kappa_2}{d\tau} &= \frac{2 (3 m^2 + 6 m + 4)}{(m+2)^3 e^m} -
\frac{2(m^4 + 8 m^3 + 30 m^2 + 68 m + 16)}{(m+2)^5 e^m} \kappa_2
}
\end{equation}
for exponential $g(w)$. 
Since we are interested in the asymptotic behavior, the above equations are
approximated for large $m$ by 
\begin{equation}
\frac{dm}{d\tau} \simeq \frac{4}{m e^m}( 1 + \kappa_2) ,\quad
\frac{d\kappa_2}{d\tau} \simeq \frac{2}{m e^m}( 3 -   \kappa_2).
\label{Eq:RLP_2}
\end{equation}
Thus we see that $\kappa_2 \rightarrow 3$ and the mean field equation for $m$ 
receives a multiplicative fluctuation correction.
By increasing $\ell$, the solution should become more accurate.

Clearly the above scheme can not be applicable to all $g(w)$. For example,
if  $g(w)$ has a power law tail \eref{powerg}, $\kappa_n$ is
infinite if $n  \ge \{ \alpha \}$. However,  
even if $\kappa_n$ is well defined for
all $n$, it is not at all guaranteed that the solution for $\langle 
w \rangle$ becomes better as we take more and more $\kappa_n$ into
account.
To clarify this point, let us think about the record problem for the exponential
distribution (\ref{expdist})
which corresponds to $\pi(x) = \Theta(x)$ and is exactly solvable.
For this case \eref{Eq:meanfit_master} and \eref{Eq:meancor_master} yield
\begin{eqnarray}
\frac{d m}{d\tau} &=& \langle e^{-w} \rangle = e^{-m} \langle e^{-\delta w} \rangle
\sim e^{-m}  \sum_{k=0}^\infty (-1)^k \kappa_k ,
\label{Eq:rec_app}\\
\nonumber
\frac{d \kappa_n}{d \tau} &=& \left \langle e^{-w} \left [ \left ( \sum_{r=0}^{n-1} 
\frac{(\delta w)^r}{r!} \right ) - \kappa_{n-1} \right ]\right \rangle\\
&\sim& e^{-m} \left ( 
\sum_{r=0}^{n-1} \sum_{k=0}^\infty  {k+r \choose k} (-1)^k \kappa_{r+k}
- \kappa_{n-1} \langle e^{-\delta w} \rangle \right )
\label{Eq:rec_app2}
\end{eqnarray}
where $\kappa_0 = 1$ and $\kappa_1 = 0$ by definition and 
$\sim$ means that the order of summation and integration 
is interchanged (without any legitimation).
The approximation scheme described above implies that
we keep terms only up to $\ell$ in (\ref{Eq:rec_app},\ref{Eq:rec_app2}).
Now assume that this solution becomes exact as $\ell \rightarrow \infty$.
If this is true, we expect that $\kappa_k \rightarrow M_k/k!$ as $\tau \rightarrow \infty$
where $M_k$ is defined in \eref{Eq:rec_cumul}.
Naively interchanging the order of summation and integration, we then get
\begin{equation}
\langle e^{-\delta w} \rangle \sim \sum_{k=0}^\infty (-1)^k \frac{M_k}{k!}\sim
 \int_0^\infty \sum_{k=0}^\infty \frac{(\gamma+\ln x)^k}{k!} e^{-x} dx
= e^\gamma,
\label{Eq:rec_gamma}
\end{equation}
which along with \eref{Eq:rec_app} gives the exact asymptotic behavior
$m \sim \ln \tau + \gamma$. 
Similarly we obtain by commuting summation and integration that 
\begin{equation}
\sum_{r=0}^{n-1} \sum_{k=0}^\infty  {k+r \choose k} (-1)^k \kappa_{r+k}
\sim e^\gamma \sum_{r=0}^{n-1} I_r = \kappa_{n-1} e^\gamma,
\end{equation}
where 
\begin{equation}
I_r = \frac{1}{r!} \int_0^\infty ( - \gamma - \ln x)^r
x e^{-x} dx = \kappa_r - \kappa_{r-1}
\end{equation}
for $k\ge 1$ and $I_0 = 1$.
Thus it seems that our approximation scheme solves the problem accurately as
$\ell$ increases.

However, this agreement is in fact fortuitous. 
\Eref{Eq:rec_gamma} illustrates the problem. One can easily see that
$\langle e^{-\delta w} \rangle$ is indeed equal to the integral yielding
$e^\gamma$, but the intermediate series 
$S_\ell \equiv \sum_{k=0}^\ell (-1)^k \kappa_k$ is actually not
convergent. In fact, even if we use the exact values for
$\kappa_n$, $S_\ell$ oscillates between 
1.500~34 and 2.0618 whose average is 1.781~07 $=e^\gamma$.
We conclude, therefore, that the approximation scheme can at best be 
expected to yield an asymptotic series.

\begin{figure}[t]
\includegraphics[width=\textwidth]{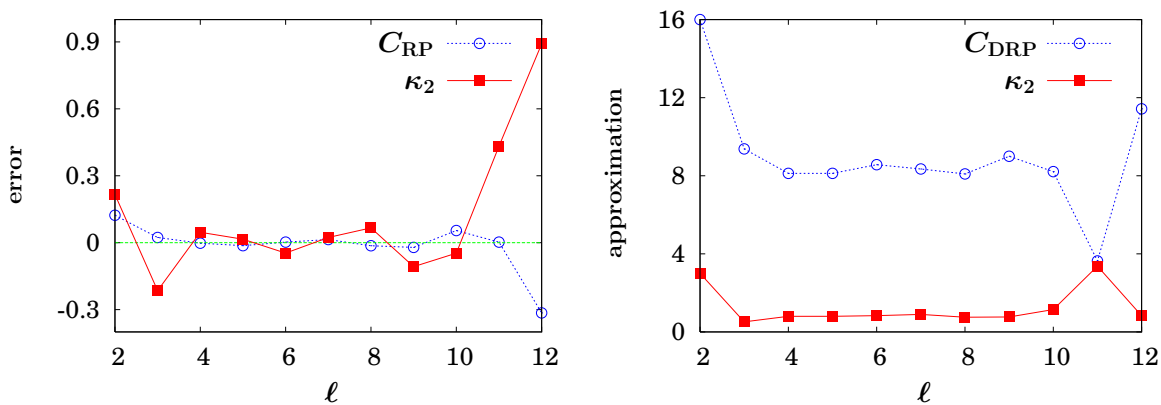}
\caption{\label{Fig:asym_app} The results of the approximation scheme for the mean 
($C_\mathrm{RP}$ and $C_\mathrm{DRP}$) and
variance ($\kappa_2$) including terms up to $\kappa_\ell$ for the RP (left panel) and for
the DRP (right panel). Since we know the exact value of $C_{\mathrm{RP}}$ and $\kappa_2$ for
the record problem, in the left panel 
we compare the approximation to the exact values.}
\end{figure}

Despite this difficulty, we now demonstrate that the moment expansion
yields reliable results when used with caution. 
We have applied the scheme to the record problem and the DRP with
exponential $g(w)$.
Assuming that all $\kappa_n$ will saturate as $\tau\rightarrow \infty$, 
the mean fitness then follows an equation of the form
\begin{equation}
\frac{dm}{d\tau} \approx C_\mathrm{RP}(\ell) e^{-m},
\end{equation}
for the record process  
\begin{equation}
\frac{dm}{d\tau} \approx \frac{C_\mathrm{DRP}(\ell)}{m e^{m}},
\label{Eq:fluc_cor}
\end{equation}
for the DRP, respectively. The argument 
$\ell$ means that the constants $C_\mathrm{RP}$ and $C_\mathrm{DRP}$ 
are evaluated keeping the $\kappa_n$ up to $n=\ell$.
For instance, we have already shown that $C_\mathrm{DRP}(1) = 4$ from 
\eref{Eq:mean_exp} and $C_\mathrm{DRP}(2) = 16$ from \eref{Eq:RLP_2}.

\Fref{Fig:asym_app} summarizes the results of the approximate evaluation
of the $C$'s and of $\kappa_2$ with increasing $\ell$.
In the range  $4 \le \ell\le 10$, the approximation  yields rather stable values.
The comparison with the exact results for the record 
problem shows that the method is excellent in this range. However,
for $\ell$ larger than 10, the error becomes uncontrollable.
In fact, the solutions for $\ell >10$ in \fref{Fig:asym_app} 
are meaningless because some $\kappa_{2 n}$'s are
negative which should be positive.
The approximation suggests that $C_\mathrm{DRP} \approx 8$, which gives a highly 
accurate estimate of the asymptotic fitness, see \fref{Fig:comall}. On the other hand,
the estimate $\kappa_2 \approx 0.8$ appears to be somewhat smaller than the simulation results, see
the inset of the right panel in \fref{Fig:WFS}.

\subsection{\label{Sec:finiteU}Finite $U$}
\begin{figure}[t]
\centerline{
\includegraphics[width=0.7\textwidth]{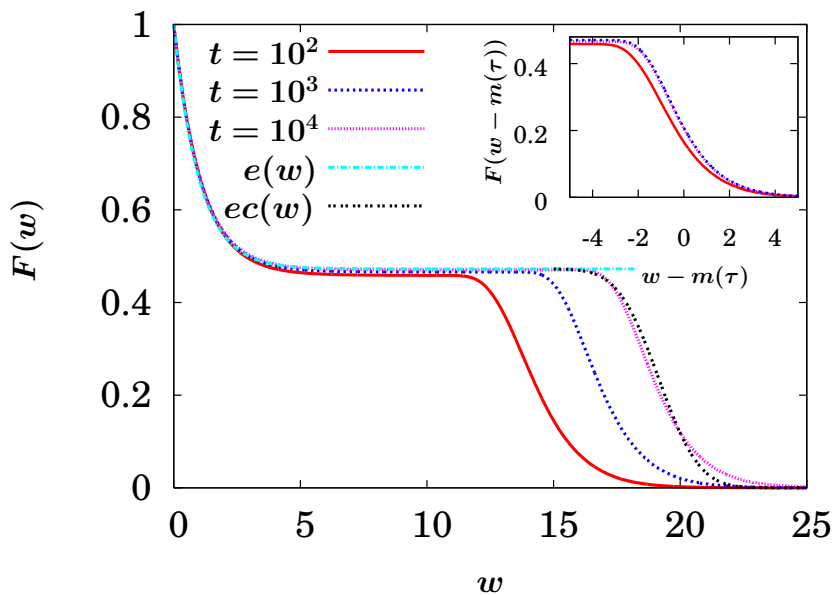}
}
\caption{\label{Fig:Wdist} The cumulative frequency distribution for $N=10^5$ and 
$U=0.5$ at $t=10^2$, $10^3$, and $10^4$ (from left to right).
The data are collected from $64~000$ independent simulations.
The approximate expressions $e(w)$ 
in \eref{Eq:Cumul_ini} and $ec(w)$ in \eref{Eq:Cumul_fi} with $\xi = 0.4725$ 
are also drawn for comparison.
Inset: same but the abscissa is shifted by the amount $m(\tau)$ which is 
the solution of \eref{Eq:fluc_cor} with $C_{DRP} = 8$. The data sets for
$t=10^3$ and $t=10^4$ show a nice collapse.
}
\end{figure}
Until now, the mutation probability $U$ has been assumed to be very small.
The natural extension of the
previous study is to ask what will happen if the 
mutation rate is high, which is the topic of this subsection. 

One can get some insight from the infinite population 
calculation in \sref{Sec:infiniteN}, where it was shown that the frequency distribution 
can be approximated by a superposition (\ref{decomp}) of two distributions 
which are well-separated from each other.
This is still expected to happen in the finite population case
when the mean fitness is much larger than the average fitness due to
mutations [the average of $g(w)$]. \Fref{Fig:Wdist} depicts the cumulative 
frequency distribution obtained from simulations of the WF model
at $t=10^2$, $10^3$, and $10^4$ for $N=10^5$ and $U=0.5$, defined as
\begin{equation}
F(w) \equiv \left \langle \int_w^\infty f_t(x;N) dx \right \rangle,
\end{equation}
where $\langle \ldots \rangle$ means an average over independent samples and
$f_t(w;N)$ is the frequency distribution for population size $N$.
As expected, the cumulative distribution displays a plateau corresponding to 
the region of low probability between the two peaks, but in contrast to the
ansatz (\ref{decomp}) the height of the plateau is below $U$. 
To get a quantitative explanation of this effect, we assume
that the frequency distribution is of the form
$f_t(w;N) = (1-\xi) g(w) + \xi \delta(w-m)$,
where $\xi$ is the weight of the high fitness peak which is approximated
by a $\delta$-function when $m \gg 1$.  
Then the population fraction
of the genotype with fitness $m$ increases after selection to $\xi m/(\xi m + 1 - \xi)$,
out of which only a fraction $1-U$ remains after mutation. If $m$ and $\xi$ 
change slowly on the time scale of one generation, one obtains the 
stationarity condition
\begin{equation}
\xi = \frac{(1-U) \xi m}{\xi m + 1 - \xi} \rightarrow
\xi = (1-U) - \frac{U}{m-1},
\end{equation}
which shows that $\xi$ approaches $1-U$ only for $m \to \infty$. 
This suggests that the cumulative distribution should take the form
\begin{equation}
F(w) \approx e(w) \equiv (1-\xi) g(w) + \xi ,
\label{Eq:Cumul_ini}
\end{equation}
for $w < m$.
For the case considered in \fref{Fig:Wdist}, the mean fitness at $t=10^4$ is
$\approx 19.15$ which gives $\xi \approx 0.4725$ in good agreement
with the simulation data. Due to 
the logarithmic increase of $m$ with $t$ in the case of exponential $g(w)$, 
the approach to the asymptotic value $\xi \to 1 - U$ is very slow.
For example, to reach $\xi = 0.49$ requires to simulate
$t \sim 10^{16}$ generations for our parameters.

\begin{figure}[t]
\centerline{
\includegraphics[width=0.7\textwidth]{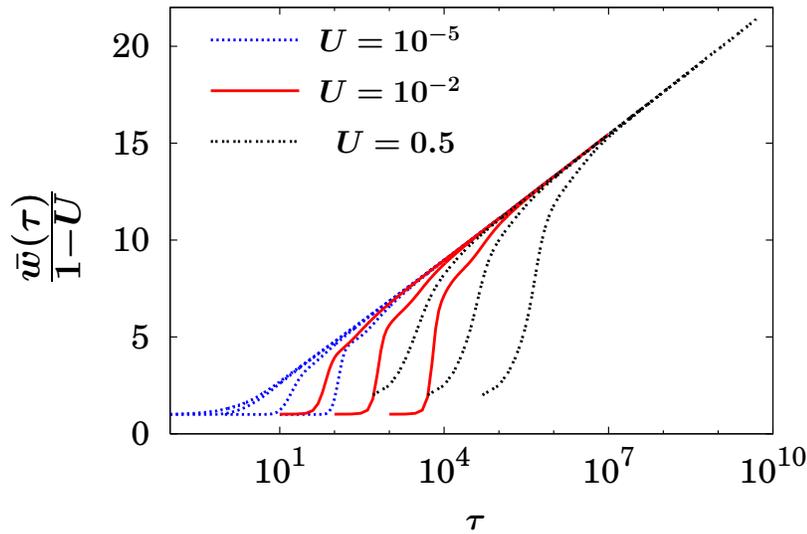}
}
\caption{\label{Fig:finiteUcollapse} Plots of $\bar w(\tau;U)/(1-U)$ vs $\tau$ for
the data sets in the left panel of \fref{Fig:WFS}. 
All curves now collapse in the asymptotic regime.}
\end{figure}
For a more accurate description of the high-fitness part of the frequency distribution
we assume that, 
as in the infinite population case, the travelling wave contribution
in the decomposition (\ref{decomp}) becomes
Gaussian at long times. If this is true, the cumulative 
frequency distribution should take the form
\begin{equation}
F(w) \approx ec(w) \equiv \frac{\xi}{2} \erfc \left ( \frac{w - m(\tau)}{2 \sqrt{\kappa_2}}
\right ),
\label{Eq:Cumul_fi}
\end{equation}
where $\erfc$ is the complementary error function and $\kappa_2 (\approx 0.8)$ is 
obtained from the analysis of the DRP in the previous section. 
As shown in \fref{Fig:Wdist}, $ec(w)$ approximates
the numerical data quite well, and the distributions 
obtained at different times $10^3$ and $10^4$ collapse 
(inset of \fref{Fig:Wdist}). 
Though $ec(w)$ is not excellent, we would say that 
it is a reasonable approximation in the asymptotic regime.

As a consequence of these considerations, the effect of the mutational load on the
mean fitness is found to be remarkably simple: Asymptotically we have 
\begin{equation}
\bar w(\tau;U) \approx \xi m(\tau) + ( 1 - \xi) = (1-U) m(\tau)
\end{equation}
where $m(\tau)$ is the mean fitness of the DRP. 
This prediction is confirmed in \fref{Fig:finiteUcollapse}.

\section{\label{Sec:Conc}Conclusions}

In this paper we have explored several aspects of the evolution of an
asexually reproducing population in a random fitness landscape which
(in the sense of the NK family \cite{Kauffman1993,WW2005}) is maximally
epistatic. In contrast to previous work on the REM fitness landscape 
\cite{Franz1993,Franz1997,Jain2005,Amitrano1989,JK2007} we adopt the limit
of infinite genome length, which leads to a finite population version of 
Kingman's house of cards model \cite{Kingman1978}. 
Although the model can hardly be expected to provide a realistic
description of empirical fitness landscapes \cite{Weinreich2005,Poelwijk2007}, it serves
as a useful counterpart to the much studied non-epistatic multiplicative
landscape model and can help to develop some intuition for the generic features
of adaptation under strong epistasis. 

An example for such a generic feature is the
slowing down of the speed of adaptation compared to the multiplicative model,
which reflects the fact that the supply of beneficial mutations dwindles with
increasing fitness. This effect is well documented in evolution experiments with
microbial populations \cite{Lenski1994,deVisser1999,EL2003}; in fact,
it has been argued \cite{Sibani1998} that the data obtained by 
Lenski and Travisano \cite{Lenski1994} for 
populations of \textit{E. coli} can be quantitatively described as
a logarithmic increase in fitness, which would be consistent with our
finite population results for exponential $g(w)$.  

We have extended Kingman's analysis of the infinite population limit to include
the shape of the frequency distribution, which was found to become bimodal
at long times. The finite population dynamics was shown to reduce to a diluted
record process (DRP) for long times and small mutation probability $U \ll 1$. This representation
allows to bound the mean population fitness from above by the standard record process 
and from below by a mean field approximation to the DRP, which can be improved systematically
through a moment expansion. Although connections between record statistics and 
evolutionary processes have been suggested before 
\cite{Kauffman1987,Kauffman1993,Sibani1998,Orr2005,Krug2005,G1994}, we have here
for the first time established a precise quantitative relation of the
theory of records to one of the cornerstones of population genetics, the WF model.
Finally, using insights from the infinite population case,
we have shown that the fitness distribution at large $U$ can be understood as a superposition 
of the DRP distribution and the mutation distribution $g(w)$.

A number of important questions concerning this model are left to future work. 
For example, it would be interesting to consider the finite population dynamics
for bounded mutation distributions $g(w)$, for which the infinite population 
calculation predicts the appearance of singularities in the stationary
fitness distribution \cite{Kingman1978}. Furthermore, the temporal statistics
of the fixation events should be investigated with regard to its relation
to record dynamics, for which detailed results are available \cite{Glick1978,Arnold1998}, as well as in 
comparison to recent work on the corresponding issue for multiplicative
fitness landscapes \cite{Park2007}. 

\section*{Acknowledgements}

We are grateful to Kavita Jain for useful discussions, and acknowledge financial
support from DFG within SFB 680 \textit{Molecular basis of evolutionary innovations}.
This paper is dedicated to Thomas Nattermann on the occasion of his 60th birthday.

\section*{Appendix A: Infinite population calculation for Gumbel-type $g(w)$}

The first step is the calculation of the $t$-th moment of (\ref{Eq:Gum_Dis}),
\begin{equation}
G_t = \int w^t g(w) dw = e w_0^t \int_{1}^\infty (\ln (y))^t e^{-y} dy.
\end{equation}
To get an asymptotic expression, let us make a change of
variables $y = x t / \ln t$
\begin{equation}
\fl
\eqalign{
G_t &= \frac{e t w_0^t}{\ln t} \left ( \ln \left ( \frac{t}{\ln t} \right )
\right )^t \int_{\frac{\ln t}{t}}^\infty dx
\left ( 1 + \frac{\ln x}{ \ln \frac{t}{\ln t}} \right )^t
\exp \left ( - \frac{t}{\ln t} x \right )\\
 &= \frac{e t w_0^t}{\ln t} \left ( \ln \left ( \frac{t}{\ln t} \right )
\right )^t \int_{\frac{\ln t}{t}}^\infty dx
\exp \left ( - \frac{t}{\ln t} \left \{ x - \ln t \ln \left ( 1 + \frac{\ln x}{ \ln \frac{t}{
\ln t}} \right ) \right \} \right ).
}
\end{equation}
Since $t$ is assumed to be very large, the integral is dominated by the regime where
the terms in the curly bracket attain a minimum which is easily seen to be unique.
One can show that the contribution from the boundaries of the integral
is exponentially small and the maximal contribution comes from the region
around $x_c$ which is the solution of the equation
\begin{equation}
(1-x_c) \ln t = x_c (\ln x_c - \ln \ln t )
\Rightarrow
x_c \simeq 1 + \frac{\ln \ln t}{\ln t} + O\left ( \frac{\ln \ln t}{\ln t} \right )^2.
\end{equation}
Expanding the terms in the curly bracket around $x_c$ and
performing the Gaussian integral, we get
\begin{equation}
\fl
G_t \simeq e w_0^t \sqrt{ \frac{2 \pi t}{\ln t} } \left ( \ln \left ( \frac{t}{\ln t} \right
)
\right )^t \exp \left ( - \frac{t}{\ln t} \left \{ x_c - \ln t \ln \left ( 1 + \frac{\ln x_c}{ \ln \frac{t}{
\ln t}} \right ) \right \} \right ).
\label{Eq:Gum_app}
\end{equation}
Hence in the long run, the Gumbel-type distribution also meets the criterion 
(\ref{criterion}) and the mean fitness becomes
\begin{equation}
\bar w_t \approx  w_0 (1-U) \ln \frac{t}{\ln t} \exp \left ( \frac{\ln \ln t}{(\ln t)^2}
\left ( 1 + \frac{\left ( \frac{3}{2} \ln \ln t - 1 \right )}{\ln t} \right ) \right ),
\label{Eq:Gum_Asym}
\end{equation}
which increases logarithmically for long times.

\section*{Appendix B: Record dynamics}

Here we derive an expression for the record probability distribution 
$p_\mathrm{RP}(w,\tau)$ of the population fitness at scaled time $\tau$,
in the long time limit where the generation of new random variables is 
equivalent to a Poisson process in continuous time \cite{Davidsen2007}.  
During time $\tau$, the probability that there are $n$ mutations is
$e^{-\tau} \tau^n/n!$. 
Hence the probability  for the fitness at time $\tau$ to be larger than $x$ is
\begin{equation}
\label{Prob>}
\mathrm{Prob}[w > x, \tau] = 
\sum_{n=0}^\infty e^{-\tau} \frac{\tau^n}{n!} \left ( 
1 - ( 1 - Q(x))^n \right )
= 1 - \exp(-\tau Q(x) ),
\end{equation}
where $Q(x) = \int_x^\infty dw \; g(w)$ is the cumulative mutation distribution and
$x$ is larger than the  initial value $w_0$ of the process. In addition there is a contribution
from the possibility that no new record has appeared up to time $\tau$,
\begin{equation}
\label{nonew}  
\mathrm{Prob}[w=w_0,\tau] = 
1 - \mathrm{Prob}[w>w_0,\tau]
= \exp(-\tau Q(w_0)).
\end{equation}
Together the two contributions yield 
\begin{equation}
p_\mathrm{RP}(w,\tau) = \tau g(w) e^{-\tau Q(w)} \Theta(w-w_0)+ 
e^{-\tau Q(w_0) } \delta(w-w_0),
\label{Eq:rec_gen}
\end{equation}
from which moments and cumulants can be derived. For example, for
the exponential case with $g(w) = Q(w) = e^{-w}$ we obtain
\begin{eqnarray}
\langle w \rangle_\mathrm{RP} - w_0
e^{-\tau e^{-w_0}} = \int_{w_0}^\infty dw \; w \tau e^{-w} e^{-\tau e^{-w} } 
= \ln \tau   +  \gamma  + O(e^{-\tau e^{-w_0}}),\\
\langle ( w - \langle w \rangle_\mathrm{RP} )^n \rangle_\mathrm{RP} =
\int_0^\infty (-\gamma - \ln x)^n e^{-x} dx + O(e^{-\tau e^{-w_0}})
\rightarrow M_n,
\label{Eq:rec_cumul}
\end{eqnarray}
where $\gamma(\approx 0.5772)$ is the Euler number. Hence 
\begin{equation}
\label{expo}
\langle w \rangle_\mathrm{RP} \approx \ln \tau + \gamma, \;\;\;
\langle (w - \langle w \rangle_\mathrm{RP} )^2
\rangle_\mathrm{RP} \approx M_2 = \frac{\pi^2}{6}, 
\end{equation}
which is consistent with extreme value statistics~\cite{D1970}.
For a Gaussian $g(w) = 2 e^{-w^2}/\sqrt{\pi}$, one finds
\begin{equation}
\fl
\eqalign{
\langle w^n \rangle_\mathrm{RP} &\approx \tau \int_0^{\erfc(w_0)} ( \erfc^{-1}(y) )^n e^{-\tau
y} dy \approx \int_0^\infty \left ( \ln (c \tau) - \ln x \right )^{n/2} e^{-x} dx\\
&\approx (\ln(c \tau))^{\frac{n}{2}} +\frac{n}{2} (\ln (c \tau))^{\frac{n}{2}-1} \gamma + \frac{n^2 - 2n}{8} 
(\ln (c \tau))^{\frac{n}{2}-2} \left ( \gamma^2 + \frac{\pi^2}{6} \right ) ,
}
\end{equation}
where $\erfc$ is the complementary error function $\erfc(x) = \frac{2}{\sqrt{\pi}}
\int_x^\infty e^{-y^2} dy$ and $\erfc^{-1}(y)$ is its inverse function whose
leading asymptotic behavior is $(\ln c-\ln y)^{1/2}$ with $c = \sqrt{2/\pi}$  when $y \ll 1$.
Hence, for the Gaussian case, 
\begin{equation}
\langle w \rangle_\mathrm{RP} \approx \ln^{1/2} (c \tau),\quad
\langle (w - \langle w \rangle_\mathrm{RP} )^2
\rangle_\mathrm{RP} \approx \frac{\pi^2}{24 \ln (c \tau)}.
\end{equation}
For a distribution with a power law tail \eref{powerg},
 we have (for $n< \alpha $)
\begin{equation}
\label{Eq:power_rec}
\langle w^n \rangle_\mathrm{RP} \approx \int_{w_0}^\infty dw \tau \alpha w^{n} (1+w)^{-\alpha - 1}
e^{-\tau (1+w)^{-\alpha}}
\approx 
\tau^{n/\alpha} \Gamma\left( \frac{\alpha - n}{\alpha} \right ),
\end{equation}
which yields (if both exist)
\begin{equation}
\label{power}
\langle w \rangle_\mathrm{RP} \sim \sqrt{\langle (w - \langle w \rangle_\mathrm{RP} )^2 
\rangle_\mathrm{RP}} \sim \tau^{1/\alpha}.
\end{equation}
If $n\ge \alpha$ in \eref{Eq:power_rec}, $\langle w^n \rangle_\mathrm{RP} $ is infinite.

\section*{References}
\bibliographystyle{unsrt}
\bibliography{me.bib,book.bib}

\end{document}